# Behavioural Dimensions for Discovering Knowledge Actor Roles Utilising Enterprise Social Network Metrics

**Janine Viol**
Friedrich-Alexander-Universität Erlangen-Nürnberg
Nuremberg, Germany
Email: janine.viol@fau.de

**Rebecca Bernsmann**
Friedrich-Alexander-Universität Erlangen-Nürnberg
Nuremberg, Germany
Email: rebecca.bernsmann@fau.de

**Kai Riemer**
The University of Sydney
Sydney, Australia
Email: kai.riemer@sydney.edu.au

## Abstract

The identification of distinct user roles is an important theme in social media research. However, for Enterprise Social Networks (ESN), the use of social media within organisations, research identifying such roles is still lacking. Yet, understanding user roles, in particular regarding their knowledge contributions and communication behaviour, might usefully support companies in managing critical knowledge resources. Against this backdrop, in this research-in-progress paper we derive 16 metrics characterising the participation behaviour, message content and structural position of ESN users of an Australian professional services firm. Based on a factor analysis, we identify four distinct dimensions of ESN user behaviour: Contribution & networking, information provision, contact dispersion and invisible usage. With this research we contribute to the literature by transferring concepts and methods of organisation science and social media research to an ESN context. Further, our approach forms the basis for the identification of different types of knowledge actors, which might ultimately help to improve organisational knowledge transparency.

## 1 Introduction

Knowledge-intensive work is often conducted within informal organisational structures (Brown and Duguid 2001b; Allen et al. 2007) that sit alongside the organisational chart and formal work procedures. As employees draw on their personal relationships to collaborate and tackle work-related problems, much of an organisation's knowledge is embedded in informal structures, which are neither fully known to a company's management nor to the individuals involved. While often highly effective, such practices raise certain managerial questions regarding the identification and development of critical knowledge resources.

Besides traditional methods, such as knowledge mapping or process analysis (Eppler 2001; Lehner 2014), approaches drawing on social network analysis promise to help identifying informal social networks by considering advice seeking and trust relationships among employees (Krackhardt and Hanson 1993). Yet, these methods involve significant manual effort, are often unreliable due to employing survey data and are not applicable to large populations (Fischbach et al. 2008). As a result, knowledge identification remains a challenging task to the extent that many organisations 'don't know what they know'.

In recent years, companies have started to use so-called Enterprise Social Networks (ESN) in order to improve knowledge sharing and collaboration among employees (McAfee 2006), well-known examples of which include IBM Connections, Jive or Yammer (Gartner 2013). ESN provide features such as user profiles, message streams, search options and different communication channels (Koch et al. 2007; Richter 2010). Compared to public and open social networking sites, such as Facebook or LinkedIn, ESN are closed applications only accessible to a company's employees.

Prior ESN research identified use practices such as knowledge creation, information sharing and problem solving (Richter and Riemer 2013; Thom et al. 2011) and points to a high relevance of ESN in the context of knowledge-intensive work (Riemer and Scifleet 2012; Schneckenberg 2009). Engaging in these practices, users leave a number of digital traces accumulated in the ESN back end (Behrendt et al. 2014).

It is the overarching goal of this research project to analyse ESN data to support knowledge identification. Accordingly, the sum of the digital traces left by users on an ESN platform are considered as a web of relations that constitutes a subset of the knowledge flows of an organisation. Against this backdrop, our research aims to unearth from ESN data different roles of ESN actors that drive the knowledge flows of an organisation, and thus act as proxies for the organisational knowledge embedded in these flows.

By drawing on studies in organisation science and online social media, in this research-in-progress paper we develop 16 metrics to characterise the behaviour of users of an Australian professional services firm's ESN. Addressing the need to cover multiple dimensions when inferring roles (Gleave et al. 2009; Trier and Richter 2014; Forestier and Stavrianou 2012; Probst et al. 2013), we consider those measures that reflect user participation, characterise the content of message posts and describe a user's position within the ESN network. By employing factor analysis we further reduce the 16 metrics to four dimensions that describe ESN user behaviour. We label the four dimensions as *contribution & networking*, *information provision*, *contact dispersion* and *invisible usage*. In a future step, a cluster analysis will aim to discover behavioural patterns that indicate ESN knowledge actor roles.

The topic of social roles has already received a lot of attention with regards to online social media in general but is not yet sufficiently studied in an ESN context (Trier and Richter 2014; Berger et al. 2014). Even though the features and design of ESN bear a lot of resemblance to their public counterparts, the user roles that emerge in a professional setting are likely to differ from the ones identified in public forums and social media. While some research is available on ESN use practices (e. g. Riemer et al. 2012; Riemer and Scifleet 2012), little is known about the extent to which users engage in different use practices and how combinations of use practices contribute to behavioural patterns and different user roles. Moreover, prior work has mostly relied on qualitative analysis of message content. To enable the analysis of larger datasets, there is a need to characterise ESN interactions quantitatively.

We contribute to the emerging research stream of ESN data analytics by laying groundwork for role discovery in ESN settings. A better understanding of ESN knowledge actor roles can help assess and improve the health of an ESN community. Moreover, while ESN reinforce collaboration and knowledge sharing, their potential to support knowledge identification has not yet been investigated. Determining knowledge actor roles in an organisation is an essential step in identifying knowledge loss risks and to create strategies for their mitigation.

The remainder of the paper is organised as follows: The next section outlines the perspective on knowledge as applied in this study and introduces the concept of organisational social networks. It further carries out an analysis of related work regarding the identification of roles in offline and online settings. The result is an overview of metrics used to identify roles in organisations and web-based applications, such as forums and ESN. Section three describes the ESN dataset to be analysed in this study. Section four provides an overview of the metrics developed so far and presents the results of the factor analysis. The final sections discuss the results of the preliminary data analysis, and outline contributions and next steps of this research.

## 2   Background and Related Work

Knowledge management research distinguishes two perspectives on organisational knowledge (Cook and Brown 1999). One stream of research considers knowledge as an object, which belongs to an individual knower (Nonaka 1994). Accordingly, knowledge management tasks focus on the extraction of individual knowledge in order to make it accessible to others in the organisation. A second stream of research assumes a knowledge-in-practice perspective which puts an emphasis on 'knowing' in that knowledge is always embedded and enacted in the organisational practice (Brown and Duguid 2001a; Orlikowski 2002). Individuals are considered as mediators of knowledge within this practice. While the knowledge-as-an-object stream suggests knowledge to be manageable like any other physical asset, the knowledge-in-practice perspective places strong emphasis on the social context of knowing and so-called knowledge work. Hence, the focus is on managing knowledge work rather than on managing knowledge as such (Newell et al. 2009). This includes the building of communities and networks to support knowledge sharing and translation (Brown and Duguid 2001a; Orlikowski 2002).

In this study we follow the knowledge-in-practice view. Accordingly, this study requires a knowledge identification method that unveils knowledge and the associated individuals by considering the social context employees interact in. As such, organisations can be considered as a web of formal and informal linkages between employees. While formal social networks are typically prescribed by management and represent the organisational structure of a corporation, informal social networks emerge from social interactions (Allen et al. 2007). Interactions on ESN thus add another layer to the web of organisational linkages.

Organisation science researchers have sought to identify different types of actors in "offline" settings, mainly considering the structural embeddedness of individuals in using concepts and methods related to Social Network Analysis (SNA) (e.g., Parise et al. 2006). On the other hand, social roles have been studied in online settings also. In this regard, social roles can be characterised by their "structural signature", which includes an individual's structural position in the online community as well as behavioural patterns (Gleave et al. 2009). In the following, we provide background information on SNA because it is the key method to characterise an individual's structural network position. We then provide an overview of selected approaches and methods used to identify different types of actors in both organisational "offline" and online settings. Finally, we introduce a set of relevant metrics to describe behaviours of knowledge actors in ESN.

## 2.1 Social Network Analysis

A social network "consists of a finite set or sets of actors and the relation or relations defined on them" (Wasserman and Faust 1994, p. 20). The transfer of immaterial or material resources, such as information and knowledge between nodes, i.e. actors, happens along relational ties (called edges) that can be directed or undirected. In a directed network an arrow pointing from person A to person B would mean that person A asked person B for advice. Undirected relationships between employees, that is a line without arrows, might show that A and B are associated, e.g. they work in the same department.

Social network analysis (SNA) offers readily developed measures to characterise formal and informal networks. The most common measures to characterise individual nodes in the network include degree centrality, betweenness centrality and closeness centrality (Freeman 1978). Degree centrality considers the number of connections of an individual actor with in-degree and out-degree determining the number of incoming or outgoing links in directed networks. Betweenness centrality measures the number of times a node acts as a bridge along the shortest path between two other nodes and closeness centrality focuses on how close an actor is to all other actors in the network.

## 2.2 Approaches that Identify Key Roles in Organisations Generally

In the past decade, a number of studies have analysed informal social networks in organisations, e.g. investigating the social aspects of knowledge sharing and development (Cross et al. 2001a; Cross et al. 2001b), forms of knowledge exchange in R&D teams (Allen et al. 2007) or detecting influential members in organisations (Cross and Prusak 2002).

All studies have in common that they calculate SNA metrics to systematically identify knowledge and knowledge flows within organisations (Chan and Liebowitz 2006). In doing so, the data to construct the network is collected from employees using questionnaires. Depending on the focus of the questionnaire, different kinds of networks can be identified, such as communication, information, or problem-solving networks (Cross et al. 2002; Toni and Nonino 2010). The relations in the resulting social network graph reflect knowledge interactions between employees. Calculating the above mentioned centrality measures, central connectors, boundary spanners, information brokers and peripheral specialists can be identified as key roles in an information network (Cross and Prusak 2002; Parise et al. 2006). In this regard, central connectors are go-to-persons due to their technical expertise that connect many individuals in an informal network. Boundary spanners and information brokers connect people from different groups and support and control the information exchange. Peripheral specialists hold specific expertise, but are peripheral in terms of their network location.

Extending organisational SNA, Helms and Buijsrogge (2005, 2006) developed an approach called knowledge network analysis. Helms and Buijsrogge (2005) differentiate between different roles such as knowledge creators, knowledge sharers and knowledge users. Visualised in a knowledge network graph, knowledge creators will have many outgoing links as they provide others with knowledge. Acting as intermediaries, knowledge sharers have both ingoing and outgoing links. Knowledge users can be recognised by having ingoing links only.

Organisational SNA as well as knowledge network analysis operationalise knowledge based on specific interactions between individuals. In this regard, behaviours such as problem solving, advice giving, advice seeking and providing knowledge are of particular interest to identify organisational knowledge flows. Individuals with prominent structural characteristics, e.g. a high in-degree centrality, are then attributed a certain role in the organisation. These roles reflect the extent to which individuals originate, distribute, develop, seek or use knowledge.

Both organisational SNA and knowledge network analysis are feasible to provide valuable insights into the knowledge flows in a company's informal network. However, the manual collection of social network data is time-consuming, may be biased and cannot be applied to analyse large datasets (Fischbach et al. 2008).

### 2.3 Approaches that Identify Roles in Online Settings

The identification of social roles in online settings has been studied in many different contexts with marketing being a major area of application (e.g., Kaiser et al. 2011). While the data collection requires less manual effort than the collection in "offline" settings, the interpretation of the network graph and the calculated metrics is a challenging task. For instance, the mere existence of a link between two users does not allow for conclusions about the quality and strength of the relationship (Gilbert and Karahalios 2009). Similarly, social roles cannot be identified by considering their structural position only (Gleave et al. 2009; Probst et al. 2013).

Recent literature reviews by Forestier and Stavrianou (2012) and Probst et al. (2013) provide an overview of how the identification of social roles and key individuals has been approached in the existing literature. Forestier and Stavrianou (2012) distinguish between the discovery of non-explicit and explicit roles. While non-explicit roles are discovered using unsupervised machine learning techniques in order to group the data into (non-predefined) categories of roles, explicit roles are defined upfront and then identified in an online social network based on a set of criteria. Focusing on the identification of influential users in their literature review, Probst et al. (2013) argue that individuals should be described based on "who one is", "what one knows", "whom one knows" and "how active one is".

Drawing on and extending the literature reviews of Forestier and Stavrianou (2012) and Probst et al. (2013), Table 1 shows the investigated platforms, focus, methods (SNA: social network analysis, AM: activity measures, CA: content analysis) and found social roles of selected articles identifying roles in online settings.

| Authors | Platform | Focus | Method | | | Social role |
|---|---|---|---|---|---|---|
| | | | **SNA** | **AM** | **CA** | |
| Welser et al. (2007) | Usenet (newsgroups) | Identification of answer people | x (replies) | x | x | Discussion person, answer person |
| Chan et al. (2010) / Angeletou et al. (2011) | Boards.ie (discussion forum) | Discovery of user roles / Identification of the users specified in Chan et al. (2010) | x (replies) | x | | Joining conversationalist, popular initiator, taciturn, supporter, elitist, popular participant, grunt, ignored |
| Hansen et al. (2010) | CSS-D (email list) | Role discovery based on network graphs | x (replies) | | | Answer people, discussion starters, questioners |
| Rowe and Alani (2012) | SAP online community (forum) | Role discovery | | x | | Focussed novice, focussed expert, participant, knowledgeable member, knowledgeable sink, initiator, mixed novice, mixed expert, distributed novice, distributed expert |
| Burns and Kotval (2013) | Engage (ESN) | Exploration of question asking and answering | x (replies) | x | x | Discussion starter, discussion replier |
| Berger et al. (2014) | Yammer (ESN) | Identification of value-adding users | x (social graph, activity graph) | x | x | Value-adding users |

*Table 1. Overview of studies identifying roles in online settings*

Depending on the investigated platform, the identified roles reflect different types of media users, communication types and knowledge roles. Most studies use a combination of social network metrics and measures characterising the participation behaviour to identify social roles. Fewer studies analyse the content of messages created by the users. For instance, Burns and Kotval (2013) distinguish questions from non-questions by scanning discussion entries in Alcatel-Lucent's ESN "Engage" for question marks and question words like "who", "what", "when" and "where".

In conclusion, there is no agreed set of roles in online communities since roles always depend on the context of the observed online community. However, behaviours such as focus dispersion, engagement, contribution, initiation, content quality, and popularity, are commonly used to characterise the roles (Rowe et al. 2013).

## 2.4 Knowledge-intensive Use Practices in ESN

ESN are highly relevant in the context of knowledge-intensive work. For instance, Riemer and Scifleet (2012) found that ESN were used (1) to build common ground based on discussions and the sharing of updates, (2) to provide input through the sharing of information, (3) to create new knowledge as well as (4) to harness existing knowledge through activities related to problem solving and advice. In building the social fabric of an organisation, ESN lay the groundwork for other knowledge work. Specifically investigating question asking on ESN, Thom et al. (2011) identified questions related to e.g. information seeking, solution seeking, people seeking, and opinion gathering.

The above studies exemplify a major part of the communication on ESN to be relevant in a knowledge management context. ESN provide an environment adapt for informal knowledge actions such as information seeking, advice giving and problem solving. Indeed, these use practices are similar to the different types of informal networks previously discovered in offline settings (cf. section 2.2).

Based on the related work, Table 2 provides an overview of metrics used to identify social roles in offline and online settings. In this connection, social network metrics, i.e. degree centrality, closeness centrality, betweenness centrality, are employed to identify roles in both offline and online settings. User activities in online settings are commonly characterised based on measures such as *number of initiated threads*, *number of replies per thread*, *number of initiated threads replied to*, *number of received replies*, and *number of users replied to*. More specific metrics consider the *number of questions asked*, the *number of questions answered* as well as the *number of points* or *likes* received by a user. To enable a comparison between different users, these metrics are often related to a reference value, for instance, *number of initial messages created / total number of initial messages*.

| | Reference | Metrics employed to characterise user behaviour |
|---|---|---|
| **Offline setting** | Welser et al. (2007) | In-degree, out-degree, out-degree centrality, power |
| | Parise et al. (2006) Parise et al. (2005) | In-degree centrality, betweenness centrality |
| | Toni and Nonino (2010) | Degree centrality, in-degree, out-degree, betweenness, closeness |
| **Online setting** | Welser et al. (2007) | Number of initiated threads, number of replies per thread, in-degree, out-degree, degree distribution in an individual's neighbourhood |
| | Chan et al. (2010) | In-degree, out-degree, reciprocity, number of posts per thread, in-degree percentage, percentage of posts replied to, percentage of threads initiated |
| | Angeletou et al. (2011) | In-degree ratio, posts replied ratio, thread initiation ratio, bi-directional threads ratio, bi-directional neighbours ratio, average posts per thread, standard deviation of posts per thread |
| | Hansen et al. (2010) | In-degree, out-degree, avgerage degree of neighbours, clustering coefficient, eigenvector |
| | Rowe and Alani (2012) Rowe et al. (2013) | Number of forums contributed to / total number of forums, number of users replied to / total number users, number of users received replies from / total number users, set of thread replies authored / total set of replies, set of thread starters authored / set of thread starters authored by all users, average points per post awarded to a user / set of posts authored |
| | Burns and Kotval (2013) | Degree, weighted degree, clustering coefficient, number of posted questions, number of replies to questions, number of questions asked / number of questions answered |
| | Berger et al. (2014) | Number of received likes, number of received bookmarks, degree centrality, closeness centrality, betweenness centrality, eigenvector centrality |

*Table 2. Overview of metrics to identify social roles in offline and online settings*

In the following, the metrics shown in Table 2 will inform the development of the metrics for the ESN data analysed in the case study. Moreover, the studies regarding knowledge-intensive use practices in ESN can inform the development of further metrics, e. g. identifying how often a user shared information or praised another individual.

## 3  Overview of the Case and Dataset

This study is carried out in cooperation with the Australian partnership of Deloitte Touche Tohmatsu. Deloitte Australia has 6,000 employees located in 14 offices in Australia and provides audit, economics, financial advisory, human capital, tax and technology services.

The ESN used by Deloitte is a browser-based platform that offers a company-wide newsfeed, allows users to create a profile, features public and private groups, the sharing of updates and files as well as communicating with others by commenting on their updates or the writing of private messages. Replies to other users' messages are displayed chronologically below the original message and grouped together in a thread-like structure. It is possible to reply to any of the previous posts in a thread at any point in time. Public groups can be viewed and joined by all network members whereas private groups are only visible to invited group members. In 2008, nine employees started to use the ESN writing 66 messages. Subsequently, the ESN attracted between 1,500-2,000 new users every year.

The dataset analysed in this research spans a one year period (1 July 2012 – 30 June 2013). It includes 61,945 messages that were posted by 3,176 users, from a total base of 6,235 registered users. Message and user-related data was exported from the backend of the ESN and provided as a .csv file. The messages file includes messages posted in the main ESN message stream, private messages as well as messages exchanged in private and public groups. Based on different IDs, the messages file allows the identification of the messages belonging to one thread, recipients of replies, and the author of each message. It also contains information as to whether a message was posted in the main stream, in a private conversation or in a public or private group and indicates if the message had an attachment. For privacy reasons, Deloitte did not provide any message contents. However, the company extracted and provided certain meta information, such as the number of words and characters in each message, tagged users or topics as well as information on the appearance of different question words, such as "how" or "why", and words indicating users praising and thanking each other, such as "well done" or "thanks". A separate dataset contains user information from on a user's profile page, such as job title and geographic location. Moreover, for a subset of users additional information was obtained from Deloitte's HR system, such as gender, nationality, and tenure.

## 4  Preliminary Data Analysis

As shown in Figure 1, our data analysis approach involves four steps.

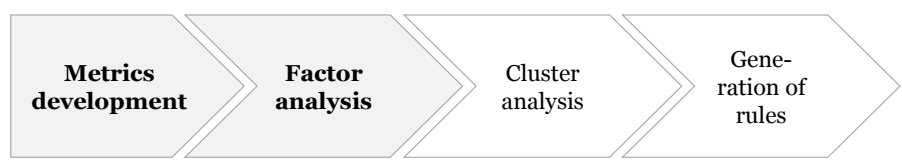

*Figure 1: Steps of the data analysis*

The first step *metrics development* includes the design of new and implementation of existing metrics from the literature (Table 2). Next, we conduct a *factor analysis* to unearth correlations among the developed metrics for Deloitte's user population, as well as to find out whether the developed metrics contribute to distinct factors, i.e. distinct behavioural dimensions. This step is expected to lead to a reduction in the set of metrics applied in the case. In the third step we will utilise *cluster analysis* to discover distinct knowledge actor roles from the case data. As such, a cluster refers to a group of users that show similar behavioural patterns. A cluster analysis is useful to discover non-explicit roles because the roles are previously unknown. Having assigned the users to different clusters, *rules* describing how the different behavioural dimensions contribute to a cluster, i.e. a knowledge actor role, can then be generated. In this paper we are able to report on the first two analysis steps.

## 4.1 Metrics Development

Based on the studies presented above and in accordance with the available messages data, we derived an initial set of metrics (Table 3). As we do not have access to message content, messages cannot be analysed in a qualitative way, for example by applying genre analysis (e.g., Richter and Riemer 2013). However, the extraction of key words such as "thanks", "well done", question words and question marks allows to make certain inferences regarding the content of messages and enables the identification of thanks messages, praise message and questions (Burns and Kotval 2013). Based on prior works, this study assumes knowledge-intensive communication to account for a major part of the overall volume of messages (e.g., Riemer and Scifleet 2012). Quantifying user behaviour on the ESN, the below metrics are thus suggested to characterise (mainly) knowledge interactions.

|  | **Metric name** | **Goal** |
|---|---|---|
| V1 | number of threads[1] contributed to / total number of threads | Determines the proportion of threads a user has contributed to |
| V2 | number of threads created / total number of threads | Measures the proportion of threads a user has initiated |
| V3 | number of threads created / number of initial messages[2] created | Determines the proportion of initial messages that yielded at least one reply |
| V4 | number of replies[3] / number of threads contributed to | Determines the average number of replies created per thread |
| V5 | number of initial messages created / number of replies created | Compares the number of initial messages with the number of replies created by a user |
| V6 | number of questions[4] created / number of initial messages created | Measures the proportion of questions created in relation to the number of initial messages created by a user |
| V7 | number of questions received / number of received tags | Determines how often a user was tagged in a question compared to the number of times a user has been tagged by other users |
| V8 | number of unique users replied to / number of replies created | Determines the proportion of unique users a user has replied to |
| V9 | number of unique users received replies from / number of replies received | Determines the proportion of unique users a user has received replies from |
| V10 | average number of words / message created | Counts the average number of words per message |
| V11 | number of attachments posted / number of messages created | Determines the proportion of a user's messages that include an attachment |
| V12 | number of received thanks[5] / number of replies received | Compares the number of thanks messages a user was tagged in with the number of replies received |
| V13 | number of received praise[6] / number of replies received | Compares the number of praise messages a user was tagged in with the number of replies received |
| V14a<br>V14b | in-degree<br>out-degree | In-degree: number of incoming links<br>Out-degree: number of links coming out of a user |
| V15 | betweenness centrality | Number of times a user acts as a "bridge" along the shortest path between two other users |

*Table 3. Metrics characterising ESN user behaviour*

All metrics shown in Table 3 refer to public messages created by individual ESN users. As such, "public" includes messages posted in the main message stream as well as messages created in public and private groups, but not private messages between users that function akin to email. Activity measures were derived using SQL statements operating on a MySQL database. All social network metrics were calculated for so-called "reply relationships" (who replied to whom), utilising the social network analysis software Gephi (Bastian et al. 2009).

---

[1] A *thread* refers to an initial message that received one or several replies.
[2] *An initial message* is a status update posted by a user that may or may not result in a thread.
[3] A *reply* is a messages directed to another message. Replies can be written to answer (undirected) initial messages as well as to comment on messages that were sent in reply to another message (reply to a reply).
[4] A message is a *question* if it is an initial message which contains a question word and a question mark.
[5] A message is a *thanks message* if it contains "thanks", "thank you" etc.
[6] A message is a *praise message* if it contains "well done", "congrats" etc.

## 4.2 Factor Analysis

In order to be considered within our analysis, users had to meet two requirements:

1) They had to be registered during the *entire study period* (1 July 2012 – 30 June 2013), i.e. 365 days.
2) Users were required to have written *at least one message per month*, i.e. a minimum of 12 messages.

The first condition ensures that all considered users had the same amount of time to contribute to the ESN. It further leads to the exclusion of employees who entered or quit the ESN during the considered year as their behaviour might be deviant and might thus distort the analysis. The second condition ensures that each user indeed actively participated in the ESN and shows a communication volume feasible to identify behavioural patterns and roles in the subsequent steps of the analysis.

Through these two criteria 682 ESN users are included in the following analyses. Factor analyses with principal axing factoring (PAF) extraction were conducted using the Statistical Package for the Social Sciences (SPSS) to group the defined metrics (Table 3). The aim of the analysis was to reduce the different metrics to a few distinct dimensions able to explain the correlation between the metrics.

PAF was initially conducted using all metrics listed in Table 3. The resulting correlation matrix shows high correlations between the metrics indicating that the variables are qualified for building factors. Also, a Kaiser's MSA value of ~0.7 verifies that the variables are well-suited for a factor analysis (Backhaus et al. 2011, p. 343). SPSS extracts as many factors as variables (Schendera 2010, p. 268), and for each factor an eigenvalue is calculated. The eigenvalue indicates how much variance of the overall variance a factor is able to explain (Schendera 2010, p. 184). A common approach is to only consider the factors with an eigenvalue greater than 1 which means that the factor explains more variance than one variable (Backhaus et al. 2011, p. 359). In our case six factors have an eigenvalue greater than 1 and together explain 71% of the variance. Metrics V1, V2, V14a, V14b and V15 can be explained by all factors, and communalities greater than 0.8 indicate that more than 80% of the variance of these variables can be explained. For metrics V7 and V10, however, only 5% of the variables' variance can be explained. Consequently, the metrics determining *how often a user was tagged in a question compared to the number of times a user has been tagged* (V7) and *the average number of words per message* (V10) do not correlate highly with the other metrics. A possible reason for V7 to not correlate might be that less than 50% of the sample was tagged in a question.

|  | Factor | | | |
| --- | --- | --- | --- | --- |
|  | 1 | 2 | 3 | 4 |
| V1 | .948 | .066 | .082 | -.095 |
| V2 | .823 | .179 | -.098 | .056 |
| V3 | .106 | -.543 | -.097 | .211 |
| V4 | .208 | -.210 | -.189 | -.089 |
| V5 | -.132 | .514 | -.140 | .095 |
| V6 | .016 | -.427 | .081 | .157 |
| V8 | -.383 | .011 | .516 | .097 |
| V9 | -.376 | .104 | .426 | .513 |
| V11 | -.066 | .535 | -.128 | .160 |
| V12 | -.182 | .067 | .366 | -.471 |
| V13 | -.121 | .025 | .271 | -.307 |
| V14a | .894 | -.025 | .206 | -.035 |
| V14b | .909 | .026 | .108 | .151 |
| V15 | .892 | .066 | .223 | .071 |

Extraction Method: Principal Axis Factoring.
(4 factors extracted; 27 iterations required)

*Table 4. Factor matrix*

As a result, the variables V7 and V10 were excluded from the analysis and a second PAF was conducted. This time four factors show an eigenvalue greater than 1 and together explain 65% of the metrics' variance. These four factors (Table 4) have been rotated to simplify the interpretation by putting each variable primarily to one of the factors (Schendera 2010, p. 249). Each variable is then assigned to the factor where it shows the highest factor loading, e.g. V1 is assigned to factor 1. To this end, a factor loading is considered "high" if it is above 0.5 (Backhaus et al. 2011, p. 362).

The most straightforward interpretation of the resulting factors can be achieved with a so-called oblique rotation (Table 5). This method allows the factors to be correlated. Within the oblique rotation, a delta of 0 (Schendera 2010, p. 246) and the Kaiser normalization were applied. The Kaiser normalization ensures that all variables have the same influence on the rotated solution (Schendera 2010, p. 211).

We are then able to interpret the resulting factors. The first factor comprises variables describing a user's quantitative contribution to and structural embeddedness in the ESN (V1, V2, V14a, V14b, V15). It is positively driven by the proportion of threads a user has contributed to (V1) and initiated (V2), as well as a user's in-degree, out-degree (V14a, V14b) and betweenness centrality (V15). The second factor is positively driven by information provision (V5, V11) and negatively driven by receiving replies to an initial message and created questions (V3, V6). The third factor mainly includes information about the variables V8 and V9, which measure the numbers of unique users a user replied to and received replies from. The fourth factor can be explained by a low manifestation of V12 and V13 that consider the number of received thanks messages and praise messages respectively. Hence, this factor is driven by low recognition. The four identified factors can thus be named (1) *contribution & networking*, (2) *information provision*, (3) *contact dispersion* and (4) *invisible usage*.

|  | Factor | | | |
|---|---|---|---|---|
|  | 1 | 2 | 3 | 4 |
| V1 | **.913** | .022 | -.130 | -.075 |
| V2 | **.751** | .167 | -.137 | .143 |
| V3 | -.033 | **-.528** | -.029 | .263 |
| V4 | .033 | -.163 | -.278 | .060 |
| V5 | -.072 | **.532** | .046 | .128 |
| V6 | -.006 | **-.448** | .109 | .104 |
| V8 | -.062 | -.101 | **.547** | -.245 |
| V9 | -.016 | -.019 | **.772** | .147 |
| V11 | .010 | **.543** | .095 | .178 |
| V12 | -.037 | .022 | .011 | **-.617** |
| V13 | -.013 | -.011 | .030 | **-.421** |
| V14a | **.919** | -.097 | .008 | -.097 |
| V14b | **.922** | -.038 | .062 | .113 |
| V15 | **.962** | -.019 | .109 | -.024 |

Extraction Method: Principal Axis Factoring.
Rotation Method: Oblimin with Kaiser Normalization.
(Rotation converged in 7 iterations)

*Table 5. Pattern matrix*

## 5　Discussion and Research Outlook

In this research-in-progress paper we have presented the initial steps of a project aiming to discover knowledge actor roles in ESN. We identified and applied 16 metrics that characterise the behaviour of users of an Australian professional services firm's ESN and reduced these metrics to four behavioural dimensions based on an exploratory factor analysis.

Our exploratory factor analysis reveals important information about ESN user behaviour. Firstly, it seems possible to discover behavioural dimensions without having to qualitatively analyse the content of messages posted to the network. Secondly, the dimension *contribution & networking* is the most important dimension because it explains 31% of the metrics' variance. Thirdly, the behavioural dimensions are correlated. Especially the factor loadings of factors 3 and 4, *contact dispersion* and *invisible usage,* change if we allow for correlation of factors (Table 5), when compared to the un-rotated factor matrix shown in Table 4 (Schendera 2010, p. 255). It appears that the dimension *invisible usage* in particular is driven by *contribution & networking*. These *invisible* users might be rather "passive" users who rarely contribute to threads and are therefore not tagged in many thanks and praise messages. Due to their low level of engagement, these users are more difficult to observe and to characterise than very active users. Since the number of "passive" users usually exceeds the number of highly active users, *invisible usage* is one topic that we plan to address in future steps.

Since we have so far only implemented a first set of metrics, the identified behavioural dimensions are preliminary. At this point, we cannot confirm, reject or extend any behavioural dimensions identified in prior work. While behaviours such as initiation, engagement and contribution were treated as distinct dimensions (e.g., Rowe et al., 2013), we find all of these aspects to be reflected in our dimension *contribution & networking*. We might hypothesise that metrics determining the level of thread initiation and thread participation correlate to a certain extent which might make it impossible to identify such distinct behavioural dimensions in an ESN setting. On the other hand, the metrics may need to be refined and extended to be able to identify more differentiated behaviour dimensions.

As a next step, we therefore aim to implement further metrics in order to characterise ESN user behaviour in more detail. To this end, we plan to develop dedicated metrics characterising who users interact with. For instance, users may or may not interact with users from other service lines, locations or levels of hierarchy. Furthermore, we aim to develop metrics that characterise user activities in groups as well as more detailed metrics regarding how users contribute to ESN conversations. For example, different users may tend to contribute at the beginning, in the middle, or towards the end of conversation threads. Being able to better show at which point they contribute could lead to the identification of more differentiated behaviours. An extended set of ESN metrics will consider a greater variety of aspects of ESN user behaviour. Employing further factor analyses, we expect to reduce the number of metrics to a final set of dimensions that will be used as input for the cluster analysis.

Finally, it would be interesting to use the proposed approach to analyse another dataset, e.g. a dataset provided by a company in another industry. The steps of our data analysis approach, e.g. the metrics development, are adaptable for different types of ESN platforms. The analysis of further datasets would allow for a comparison of the behavioural dimensions and roles identified in this study across different companies.

# 6  Conclusion

The very nature of knowledge, decentralised organisational structures and increasing information overload make it difficult for organisations to achieve knowledge transparency. The analysis of ESN data in order to support knowledge identification addresses a current and increasingly important challenge.

The contributions of our research project are threefold: Firstly, we combine concepts and methods of organisation science and social media research and transfer these to an ESN context. Grounded in these two fields, our study will improve the notion of knowledge management related user roles in ESN – and hence address a gap in the current literature (Trier and Richter 2014). For knowledge management theory, a typology of ESN knowledge actor roles enables a better understanding of knowledge processes. These concern the ways in which and by whom knowledge embedded in the organisational practice is shared, integrated, translated and transformed (Newell et al. 2009). Secondly, we contribute to the emerging field of ESN analytics by deriving metrics to quantitatively describe and analyse ESN user behaviour. Thirdly, we lay the groundwork for a novel approach to knowledge identification. While many companies are introducing or are about to introduce ESN to reinforce knowledge sharing and collaboration, we suggest the record of knowledge interactions stored in the ESN back end to be suitable to identify knowledge. Thus, the results of this research project can help run knowledge management processes, e.g. knowledge retention efforts, in a more target-oriented and sustainable way.

Beyond knowledge identification, a notion of ESN user roles may facilitate more evidence-based decision-making at the intersection between knowledge management and human resources management. At the individual level, managers could relate an employee's ESN user role with their existing "offline" performance score to identify discrepancies. For instance, individuals who create value on the ESN by frequently helping others to solve work-related problems may have less time to reach their "offline" targets and receive a below average score. If an organisation values a helpful attitude, an additional dimension, e.g. based on an employee's behaviour on the ESN, could be added to the performance management system (Cross et al. 2002; Cross et al. 2001a). At the team level, an understanding of different ESN user roles could support staffing decisions. Regarding the overall organisation, findings from ESN data analytics could help to identify and explain differences in performance between different divisions (Smith et al. 2009). Within the scope of this paper, we can only provide a few examples of how ESN data analytics can inform organisational decision-making. Further research is needed regarding particular use cases that could benefit from the derived insights.

## 7　References

## Copyright